# Correlations Between the Dielectric Properties, Domain Structure Morphology and Phase State of $Bi_{1-x}Sm_xFeO_3$ Nanoparticles

Oleksandr S. Pylypchuk[1], Vladyslav O. Kolupaiev[1], Victor V. Vainberg[1], Vladimir N. Poroshin[1], Ihor V. Fesych[2,3], Lesya D. Demchenko[4,5], Eugene A. Eliseev[6*], and Anna N. Morozovska[1†]

[1]*Institute of Physics, National Academy of Sciences of Ukraine, 46 Nauky Ave, Kyiv, 03028, Ukraine*

[2]*Taras Shevchenko National University of Kyiv, 01030 Kyiv, Ukraine*

[3]*Donetsk Institute for Physics and Engineering named after O.O. Galkin of the National Academy of Sciences of Ukraine*

[4] *Stockholm University, Department of Chemistry, Sweden*

[5] *Ye. O. Paton Institute of Materials Science and Welding, National Technical University of Ukraine "Igor Sikorsky Kyiv Polytechnic Institute", 37, Beresteisky Avenue, Kyiv, Ukraine, 03056*

[6] *Frantsevich Institute for Problems in Materials Science of the National Academy of Sciences of Ukraine, 3, str. Omeliana Pritsaka, 03142 Kyiv, Ukraine*

## Abstract

Nanoscale multiferroics are basic model objects for studying polar, magnetic and magnetoelectric properties and mutual couplings. Bismuth-samarium ferrite ($Bi_{1-x}Sm_xFeO_3$) is a model orthoferrite, whose polar, magnetic and magnetoelectric properties have been studied for the bulk and thin film samples. The properties of $Bi_{1-x}Sm_xFeO_3$ nanoparticles have been much less studied, despite the nanoparticles can be used in a wide range of applications, such as energy storage, magnetic hyperthermia and advanced nanoelectronics. In this work we performed experimental measurements and analysis of the temperature dependence of the $Bi_{1-x}Sm_xFeO_3$ nanopowders dielectric properties. Calculations of the ferro-ionic coupling influence on the dielectric properties, domain structure morphology and phase states are performed in the framework of the Ginzburg-Landau-Devonshire-Stephenson-Highland approach. Theoretical results explain the main trends of experimentally observed temperature dependences of the effective dielectric permittivity, which allows us to understand the correlations between the temperature behavior of dielectric properties, domain structure morphology and phase state of $Bi_{1-x}Sm_xFeO_3$ nanoparticles.

[*]Corresponding author, e-mail eugene.a.eliseev@gmail.com

[†]Corresponding author, e-mail anna.n.morozovska@gmail.com

## 1. Introduction

Nanoscale multiferroics, especially solid solutions of bismuth ferrite, both oxygen-deficient and doped with rare-earth ions of Sm, La, Pr, Eu [1, 2, 3], raise constant and significant scientific interest due to their interesting polar, magnetic and especially magnetoelectric properties [4, 5], which can be controlled by size effects [6, 7, 8], deformation engineering [9, 10], as well as by the level of defects, such as oxygen vacancies [11] and/or rare earth dopants [12, 13]. However, in the practical implementation of such control, it is necessary to change the technological process of the synthesis of films and nanoparticles.

Modification of the surface state of ferroelectric perovskites is possible due to the surface electrochemical reaction in perovskites [14, 15]. As a result of electrochemical reactions at the surface, the so-called “chemical” polarization switching occurs [16, 17, 18]. The surface-driven electrochemical states can significantly improve the polar properties of thin films and nanoparticles of perovskite ferroelectrics, such as $PbTiO_3$ [14-18], $BaTiO_3$ [19, 20], antiferroelectrics $PbZrO_3$ [21], silicon-compatible ferroelectric-antiferroelectric oxides $Zr_xHf_{1-x}O_2$ [22, 23], chalcogenides $Sn_2P_2S_6$ [24, 25] and $CuInP_2S_6$ [26], and multiferroics $Bi_{1-x}Sm_xFeO_3$ [27, 28]. The main reason of this improvement is the surface adsorption (or desorption) of oxygen ions, which leads to the appearance of mixed “ferro-ionic” states [29, 30, 31] in thin films and nanoparticles of the polar materials.

The properties of the rare-earth doped and/or co-doped $BiFeO_3$ nanoparticles [32, 33], fine-grained ceramics [34, 35], and nanocomposites [36] have been much less studied theoretically and experimentally compared to thin films, despite they can be used in a wide range of applications, such as energy storage [37], magnetic hyperthermia [38], and composite-based semiconductor devices.

In the previous works [27, 28] we used the Ginzburg-Landau-Devonshire approach to calculate polar and dielectric properties of $Bi_{1-x}Sm_xFeO_3$ nanoparticles, and construct their phase diagrams in dependence on the nanoparticle average size and samarium content "x". In this work we perform experimental studies of the temperature dependence of the $Bi_{1-x}Sm_xFeO_3$ nanopowders dielectric permittivity. Calculations of the ferro-ionic coupling influence on dielectric and polar properties are performed in the framework of combined Ginzburg-Landau-Devonshire (LGD) [27, 28] and Stephenson-Highland (SH) [14, 15] approaches. The main trends of experimental observations are explained by theoretical modelling, which allow us to establish correlations between the temperature behavior of dielectric properties, domain structure morphology and phase state of $Bi_{1-x}Sm_xFeO_3$ nanoparticles.

## 2. Materials and Methods

A series of nanopowder samples of $Bi_{1-x}Sm_xFeO_3$, where the Sm content "x" varies from 0 to 0.2 (shortly "BSFO"), were synthesized using the solution combustion method and then calcined for 5 hours at 750 °C to minimize the presence of residual water and hydroxyl groups. Preparation details are given in Ref. [27]. The obtained five samples were named as the BFO, BSFO-005, BSFO-010, BSFO-015 and BSFO-020, respectively, where the numbers indicate the Sm content x = 0, 0.05, 0.1, 0.15, 0.2. According to the X-ray diffraction (XRD) data [27], the BFO sample contains 73 % of the long-range ordered rhombohedral (*R*3c) $BiFeO_3$ phase, 16 % of the orthorhombic (*Pbam*) $Bi_2Fe_4O_9$ phase and 11% of the cubic (*I23*) $Bi_{25}FeO_{40}$ phase.

The Sm-doping very strongly increases the phase purity of the nanopowders. The BSFO-005 contains 97% of the $Bi_{0.95}Sm_{0.05}FeO_3$ in the *R*3c phase, 2 % of the $Bi_2Fe_4O_9$ and 1% of the $Bi_{25}FeO_{40}$. The BSFO-010 contains 99% of the $Bi_{0.9}Sm_{0.1}FeO_3$ in the *R*3c phase and 1% of the $Bi_{25}FeO_{40}$. The BSFO-015 contains 7 % and 92 % of the $Bi_{0.85}Sm_{0.15}FeO_3$ in the polar *R*3c and orthorhombic *Pbnm* phases, respectively, and 1% of the $Bi_{25}FeO_{40}$. The BSFO-020 sample contains 99 % of the $Bi_{0.8}Sm_{0.2}FeO_3$ in the *Pbnm* phase and 1% of the $Bi_{25}FeO_{40}$.

To measure temperature dependencies of the samples capacitance the $Bi_{1-x}Sm_xFeO_3$ nanopowders were pressed in the polytetrafluorethylene (PTFE) cells between two metallic plungers, which serve as electric contacts (see **Fig. 1(a)**). The samples in a cell have a disk shape with the 4 mm diameter and 0.2 mm thickness. The pressure to keep samples resistance in reasonable limits was about 2.5 MPa. The bigger pressure applied to the samples does not result in noticeable changes in their electrophysical characteristics and measured values do not change after the pressure removal.

The PTFE Cell with plungers providing a sample compression was placed inside the holder rod, which was connected to the RLC meter via coaxial cable. The RLC meter UNI-T UT612 was used to measure the capacitance of the cells in the range (100 – $10^5$) Hz. Heating of the samples was performed by the wire-wound nichrome heater connected to the controlled voltage source.

Typical transmission electron microscopy (TEM) images of the $Bi_{1-x}Sm_xFeO_3$ ($0 \le x \le 0.2$) nanoparticles are shown in **Figs. 1(b)-(g)**. From the images, the scattering of nanoparticles sizes is large enough (from 50 to 500 nm).

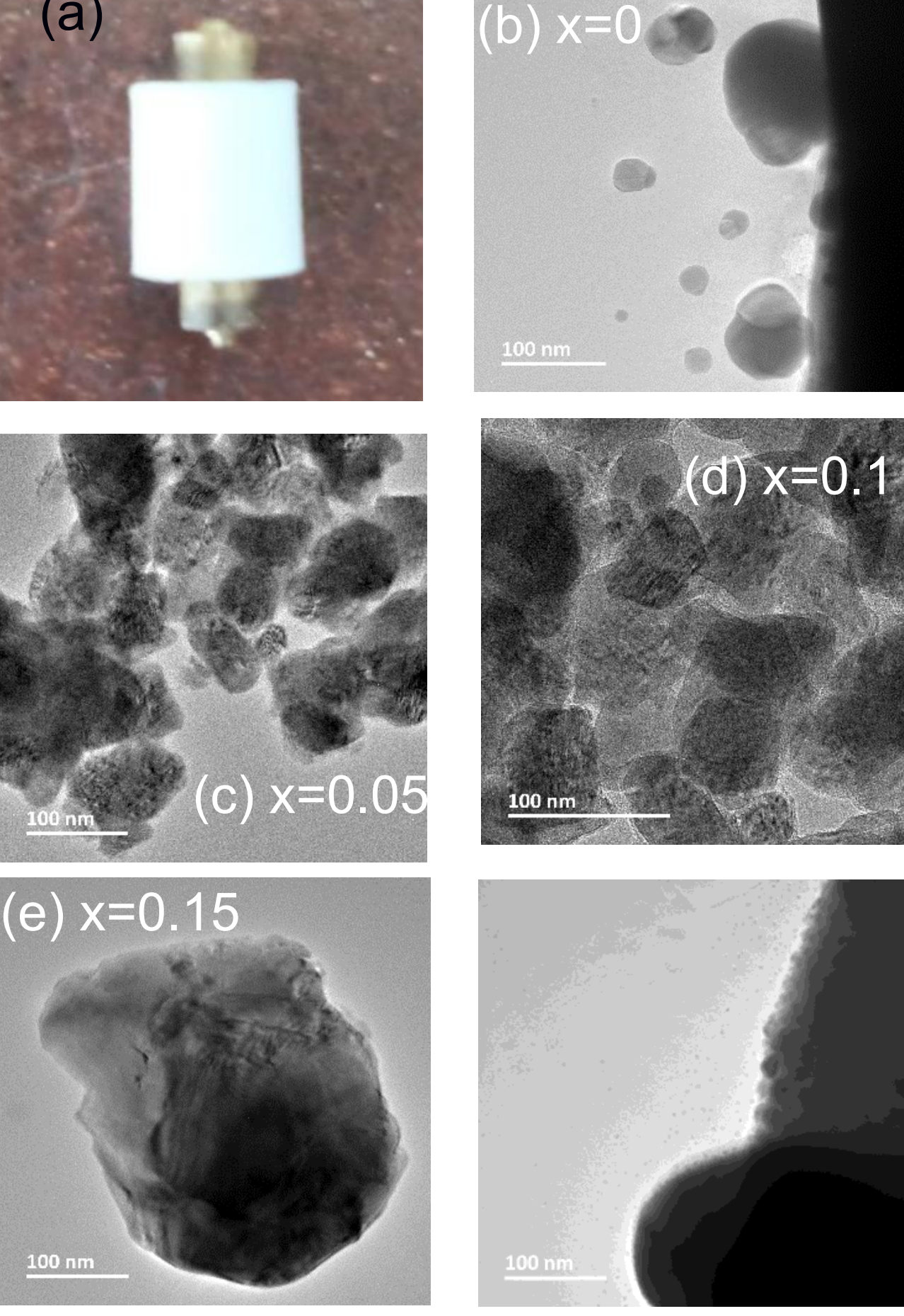

**FIGURE 1. (a)** Schematic illustration of the cell with the sample. (**b)-(g)** Typical TEM images of the $Bi_{1-x}Sm_xFeO_3$ (0≤x≤0.2) nanoparticles.

## 3. Experimental results and discussion

The temperature dependences of the capacitance of the studied samples with different content of Sm were measured at the frequencies 100 Hz and 100 kHz in the temperature range from 20 to 400 $^{o}$C. Typical results are shown in **Fig. 2**. The dependences have similar behavior for all Sm content "x" and consist of two characteristic parts. The first part, corresponding to the temperature range from 20 $^{o}$C to 250 – 300 $^{o}$C the capacitance, and consequently the effective dielectric permittivity of the nanopowders, is almost constant manifesting only very small increase in magnitude (from 10 to 60 pF) with growing temperature, which corresponds to the effective dielectric permittivity of several tens. At higher temperatures from 250 – 300 C to 400 C the capacitance of all samples manifests rapid and strong growth with growing temperature, with the tendency to shaping a maximum at the highest temperature.

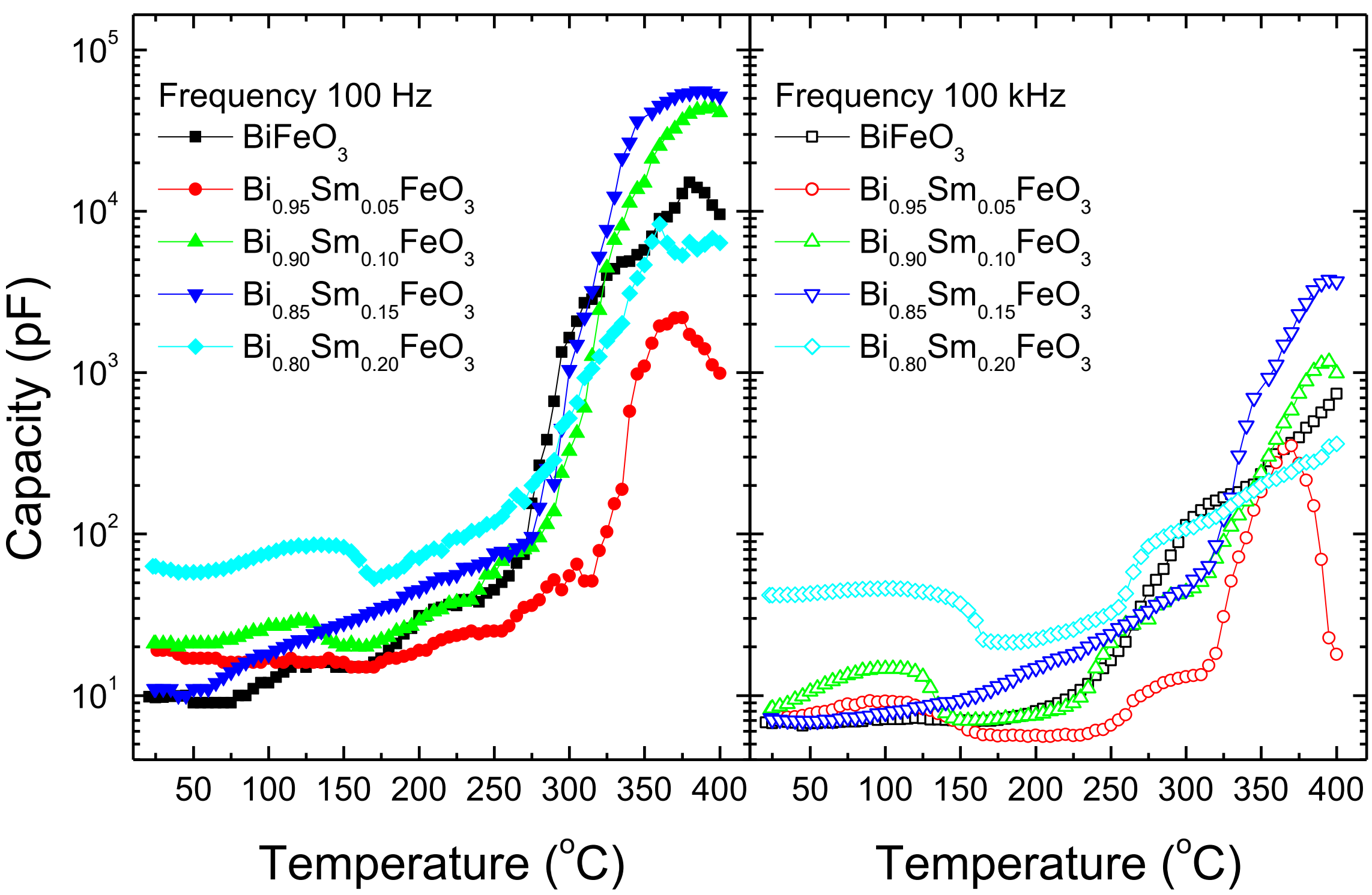

**FIGURE 2.** Temperature dependences of the pressed $Bi_{1-x}Sm_xFeO_3$ nanopowders tablets capacitance, measured at the frequencies 100 Hz and 100 kHz.

The ratio between the lowest and largest magnitudes of reative dielectric permittivity varies non-monotonously in the range from $10^2$ to $10^4$ with increasing Sm content in $Bi_{1-x}Sm_xFeO_3$ (see **Fig. 3(a)**). From **Fig. 3(a)**, the highest value of the dielectric permittivity is obtained at the Sm contents 10 % and 15 %, while at 5 % and 20 % it is more than an order of magnitude less.

Another important characteristic is the temperature of switching between the low and high dielectric permittivity regions. Shown in **Fig. 3(b)** is the dependence of temperature at which the strong growth of the dielectric permittivity begins on the Sm content. It is also non-monotonic. At first, it sharply increases with the increase in Sm content from 0 to 5 %, then it almost linearly decreases with the increase in Sm content from 5 % to 20 %.

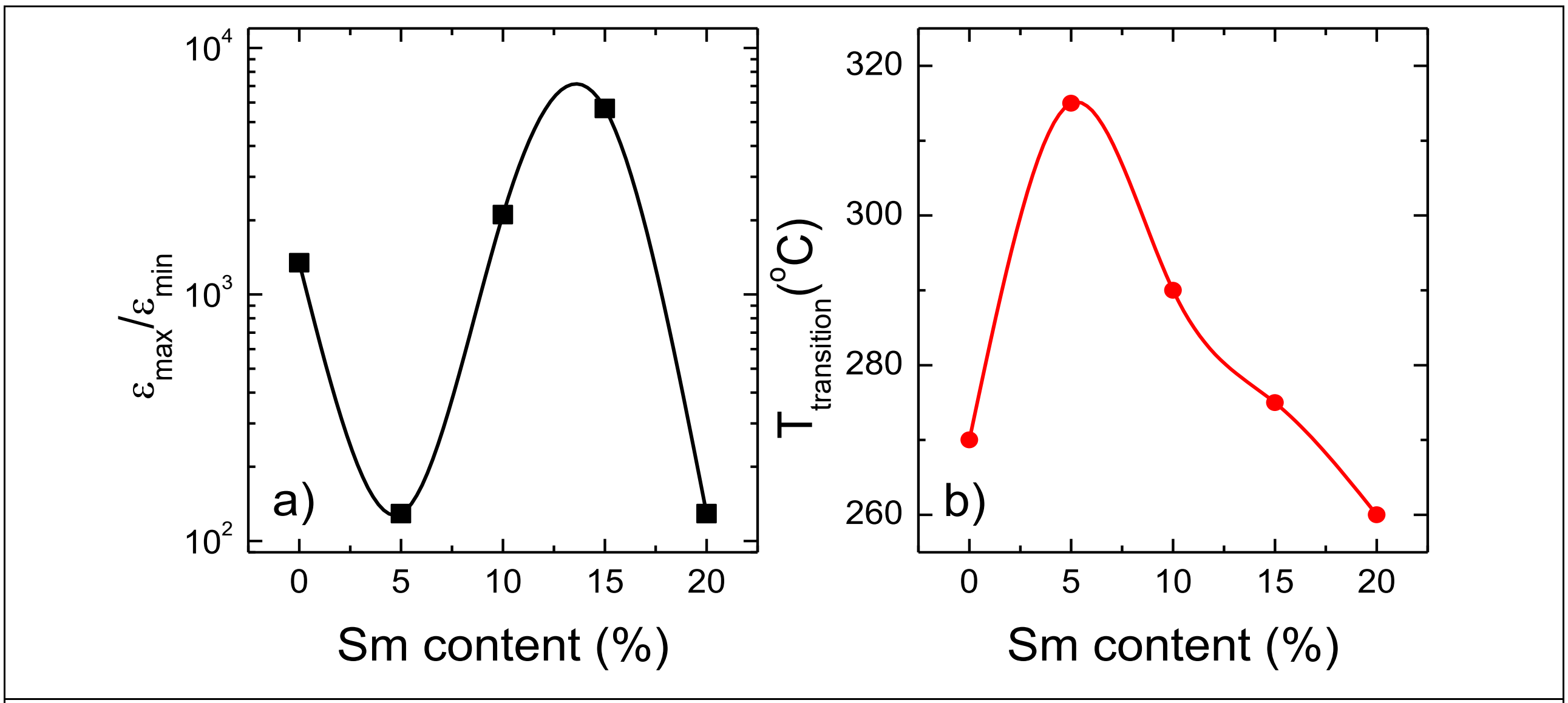

**FIGURE 3.** **(a)** The ratio between the maximal and minimal dielectric permittivity vs the Sm content in the $Bi_{1-x}Sm_xFeO_3$ nanopowders at 100 Hz. **(b)** Dependence of the transition temperature between the low and high dielectric permittivity vs the Sm content in $Bi_{1-x}Sm_xFeO_3$ nanopowders at 100 Hz.

The temperature dependences of the capacitance measured at the frequency of 100 kHz qualitatively repeat the behavior of those measured at 100 Hz. At that, the capacitance magnitude, and that of the dielectric permittivity, somewhat decreases and the temperature of the transition between the low and high dielectric permittivity regions shifts to a higher temperature by a small value, which is shown in **Fig. 4** on the example of the sample with 15 % of Sm content, which has the highest value of permittivity (as shown in **Fig. 3(a)**).

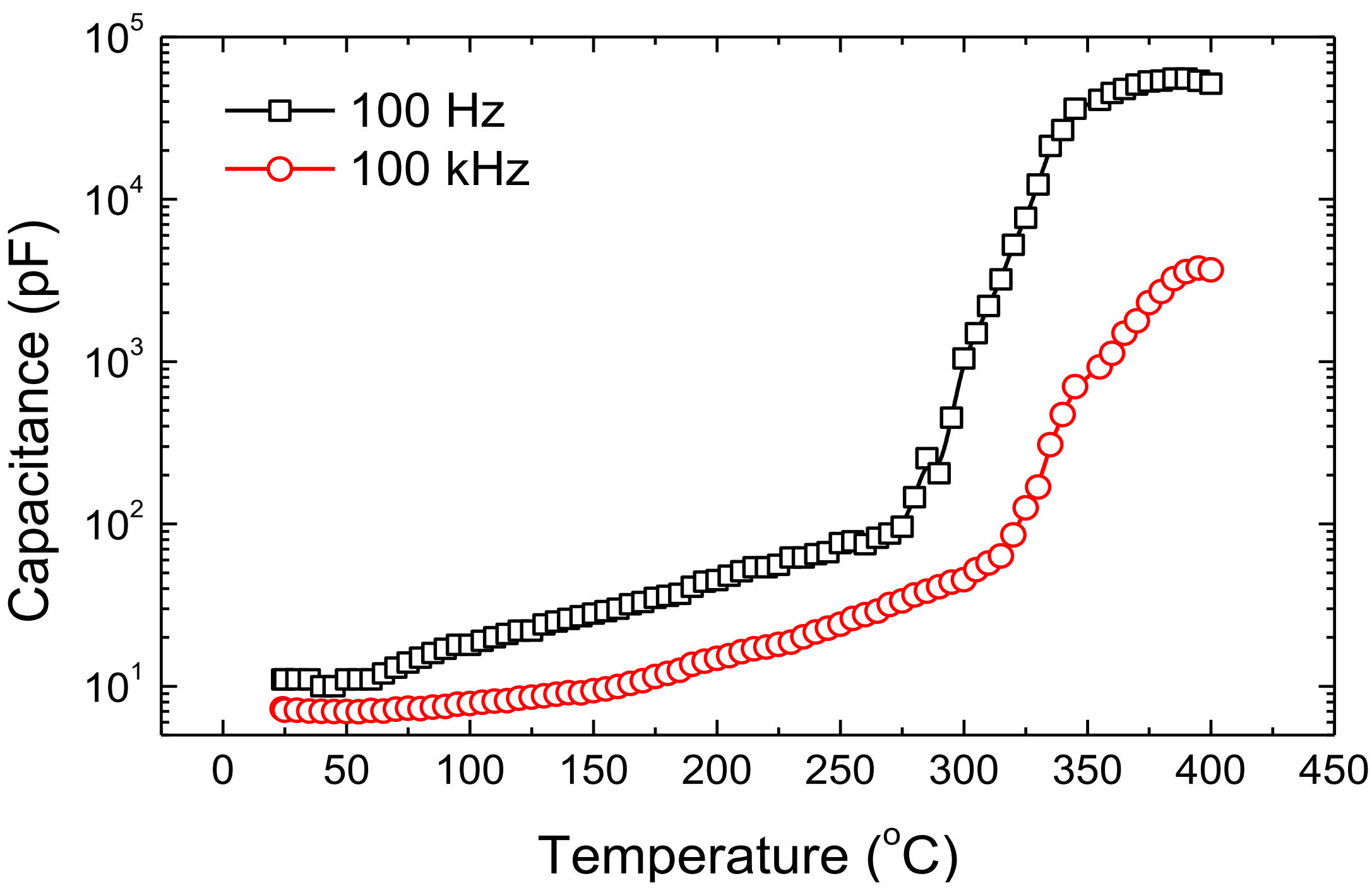

**FIGURE 4**. Comparison of capacitance temperature dependences measured at the frequencies of 100 Hz and 100 kHz on example of the $Bi_{1-x}Sm_xFeO_3$ pressed nanopowder with 15 % of Sm.

## 4. Theoretical modeling

As it was shown in the works [39, 40], that the colossal dielectric constant of heterogeneous ferroelectric ceramics [41], nanopowders [42] and/or composites is usually caused by mesoscopic inhomogeneities in electrical conductivity (mainly between grains and grain boundaries), known as the effect of internal barrier layer capacitance (IBLC), as well as by inhomogeneous layers between electrodes and the sample, known as the effect of surface barrier layer capacitance (SBLC). The IBLC effect causes a colossal dielectric constant at low frequencies if the conductivity of the substance between grains is significantly lower than the conductivity of the grains. In this case, the percolation of components with low conductivity (i.e., substances between ceramic grains) occurs and is described by the effective medium approach (EMA), discussed in Refs. [43, 44], as well as in the works of Petzelt et al. [45] and Richetsky et al. [46].

The theoretical model of four cation sublattices, proposed in Ref. [47], allows the analytical description of cation displacement in the $Bi_{1-x}Sm_xFeO_3$ compounds, at that the Sm content "$x$" and the temperature $T$ determine the stability of the long-range ordered rhombohedral ferroelectric (FE), mixed ferrielectric (FEI), antiferroelectric (AFE), and paraelectric/dielectric nonpolar (NP) orthorhombic phases. The free energy functional of the $Bi_{1-x}Sm_xFeO_3$ nanoparticles with ferro-ionic coupling was derived using the model of four sublattices, Landau-Ginsburg-Devonshire and Stephenson-Highland [14, 15] approaches in Ref. [27].

The phase state of spherical $Bi_{1-x}Sm_xFeO_3$ nanoparticles with the size 100 nm as a function of Sm content x and temperature $T$ is shown in **Fig. 5(a).** The spontaneous polar and antipolar order parameters as a function of Sm content x and temperature $T$ are shown in **Fig. 5(b)** and **5(c)**, respectively. The figures are calculated for the effective length of surface screening $\lambda$=1 nm, which originates from the ferro-ionic coupling with adsorbed layer of ambient charges (e.g., oxygen vacancies), and material parameters of $Bi_{1-x}Sm_xFeO_3$, listed in Ref. [27], except for the critical concentration of the antipolar order that is taken as $x_a$ =0.35 for better correlation with experimental results.

From **Figures 5(a)-(c)**, the spontaneous polar order exists in the FE and FEI phases. The spontaneous antipolar order exists in the FEI and AFE phases, and gradually disappears approaching the FEI-FE boundary, as well as it disappears approaching the AFE-NP boundary. The antipolar order reaches maximal values at the FEI-AFE boundary. These results are also valid for the nanoparticles with the average radius $R$ =50 nm and negligibly small dispersion of sizes.

Typical temperature changes of the ferroelectric domain structure inside the 100-nm $Bi_{1-x}Sm_xFeO_3$ nanoparticle at x = 0.08 are shown in **Fig. 5(d)**. The single-domain state at first transforms into the bi-domain with increase in temperature, then to the poly-domain, then domain stripes become low-contrast and vanish above the FE-NP phase transition temperature. Note that the temperature effects of the domain structure morphology become significant near the FEI-AFE boundary also [27].

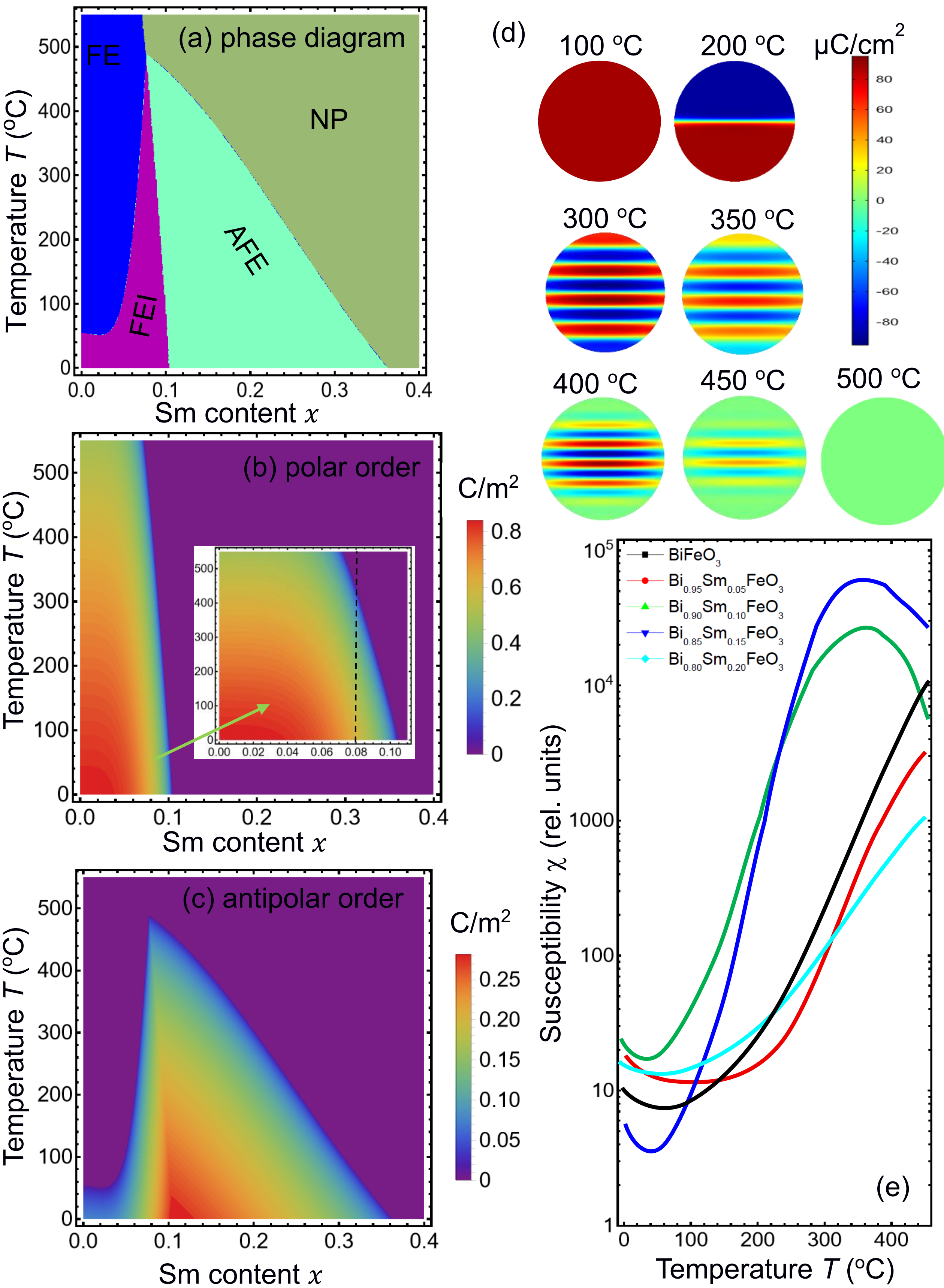

**FIGURE 5.** The phase state **(a)**, amplitudes of the spontaneous polar **(b)** and antipolar order **(c)** parameters as a function of Sm content x and temperature $T$ calculated for spherical $Bi_{1-x}Sm_xFeO_3$ nanoparticles with the size 100 nm and effective surface screening length λ=1 nm. **(d)** Temperature changes of the ferroelectric domain structure inside the 100-nm $Bi_{1-x}Sm_xFeO_3$ nanoparticle at x =

0.08. **(e)** Temperature dependences of the $Bi_{1-x}Sm_xFeO_3$ nanoparticles dielectric susceptibility (in relative units) calculated for growing Sm content x = 0 (black curve), 0.05 (red curve), 0.1 (green curve), 0.15 (blue curve) and 0.2 (cyan curves); the screening length $\lambda$=1 nm, Maxwellian distribution of sizes with the most probable radius $R$ =50 nm and dispersion 100 nm.

Temperature dependences of the $Bi_{1-x}Sm_xFeO_3$ nanopowder dielectric susceptibility (in relative units) calculated for growing Sm content x from 0 to 20 % are shown in **Fig. 5(e)** by the different curves. The dielectric susceptibility demonstrates characteristic features, such as local minimum or maximum corresponding to the FE-FEI, FEI-AFE or FE-NP phase transitions (compare different curves in **Fig. 5(e)**). The curves are smooth, because they are averaged over a relatively wide size distribution of the nanoparticles (in agreement with the TEM analysis). The sharp divergency corresponds to the second order phase transition between the FEI and AFE phases (or FE-NP phases) in the ensemble of monodispersed nanoparticles. The temperature dependences of the $Bi_{1-x}Sm_xFeO_3$ nanopowder dielectric susceptibility, shown in **Fig. 5(e)**, correlate with the physical picture, shown in **Fig. 3**, however there are significant differences, which we relate with the contribution of the Maxwell-Wagner mechanism, manifested as IBLC and SBLC effects.

## Conclusions

We performed experimental measurements and analysis of the temperature dependence of the $Bi_{1-x}Sm_xFeO_3$ nanopowders dielectric permittivity. The effective dielectric permittivity of Sm-doped bismuth ferrite manifests interesting behavior with two characteristic temperature ranges, namely a slow increase from 20 °C to 250 – 300 °C followed by a sharp increase from 300 °C to 400 °C with the tendency of maximum formation. The magnitude of the dielectric permittivity in these ranges and the transition temperature between them depend non-monotonically on the Sm content.

Calculations of the ferro-ionic coupling influence on the dielectric properties, domain structure morphology and phase states are performed in the framework of the Ginzburg-Landau-Devonshire-Stephenson-Highland approach. Theoretical results explain the main trends of experimentally observed temperature dependences of the effective dielectric permittivity, which allows us to understand the correlations between the temperature behavior of dielectric properties, domain structure morphology and phase state of $Bi_{1-x}Sm_xFeO_3$ nanoparticles.

The chemically controlled dielectric permittivity is a key for controlling the electrophysical properties, such as magnetoelectric effect and electric conduction, for example, based on the Heywang model [48]. Thus, the results of the work can help to create ferroics with improved and/or controllable polar and dielectric properties, as well as expand the perspectives of their advanced applications in nanoelectronics and energy storage.

**Authors' contribution.** O.S.P. and A.N.M. generated the research idea. O.S.P. V.O.K., V.V.V. and V.N.P. performed dielectric measurements and analyzed results. I.V.F. prepared the samples and characterized them. L.D.D. performed TEM analysis. and formulated the problem. E.A.E. performed numerical modelling and prepared corresponding figures. A.N.M. formulated the theoretical problem, performed analytical calculations and analyzed results. A.N.M., V.O.K., and V.V.V. wrote the paper. V.N.P., E.A.E. and A.N.M. coordinated the research.

**Acknowledgements.** The work of O.S.P. and A.N.M. are funded by the National Research Foundation of Ukraine (project "Manyfold-degenerated metastable states of spontaneous polarization in nanoferroics: theory, experiment and perspectives for digital nanoelectronics", grant N 2023.03/0132). V.O.K., V.V.V. and V.N.P. acknowledges the Target Program of the National Academy of Sciences of Ukraine, Project No. 5.8/25-П "Energy-saving and environmentally friendly nanoscale ferroics for the development of sensorics, nanoelectronics and spintronics". The work of E.A.E. is funded by the National Research Foundation of Ukraine (project "Silicon-compatible ferroelectric nanocomposites for electronics and sensors", grant N 2023.03/0127). I.V.F. is grateful to the II European Chemistry School for Ukrainians for providing a comprehensive overview of current trends in European chemical science. L.D. acknowledges support from the Knut and Alice Wallenberg Foundation (grant no. 2018.0237) for TEM research.

## References

[1] D. V. Karpinsky, E. A. Eliseev, Fei Xue, M. V. Silibin, A. Franz, M. D. Glinchuk, I. O. Troyanchuk, S. A. Gavrilov, V. Gopalan, Long-Qing Chen, and A. N. Morozovska. "Thermodynamic potential and phase diagram for multiferroic bismuth ferrite ($BiFeO_3$)". npj Computational Materials **3**:20 (2017); https://doi.org/10.1038/s41524-017-0021-3

[2] M. D. Glinchuk, A. N. Morozovska, D. V. Karpinsky, and M. V. Silibin. Anomalies of Phase Diagrams and Physical Properties of Antiferrodistortive Perovskite Oxides. (Author Review). Journal of Alloys and Compounds **778**, 452-479 (2019) https://doi.org/10.1016/j.jallcom.2018.11.015

[3] A. N. Morozovska, D. V. Karpinsky, D. O. Alikin, A. Abramov, E. A. Eliseev, M. D. Glinchuk, A. D. Yaremkevich, O. M. Fesenko, T. V. Tsebrienko, A. Pakalniskis, A. Kareiva, M. V. Silibin, V. V. Sidski, S. V. Kalinin, and A. L Kholkin. A Combined Theoretical and Experimental Study of the Phase Coexistence and Morphotropic Boundaries in Ferroelectric-Antiferroelectric-Antiferrodistortive Multiferroics. Acta Materialia, **212**, 116939 (2021) https://doi.org/10.1016/j.actamat.2021.116939

[4] M. Fiebig, Revival of the magnetoelectric effect. Journal of physics D: applied physics **38**, R123 (2005), https://doi.org/10.1088/0022-3727/38/8/R01

[5] M. Fiebig, T. Lottermoser, D. Meier, M. Trassin. The evolution of multiferroics. Nat Rev Mater **1**, 16046 (2016), https://doi.org/10.1038/natrevmats.2016.46

[6] M.D. Glinchuk, E.A. Eliseev, A.N. Morozovska, R. Blinc. Giant magnetoelectric effect induced by intrinsic surface stress in ferroic nanorods. Phys. Rev. B 77, № 2, 024106-1-11 (2008).

[7] R. Maran, S. Yasui, E. Eliseev, A. Morozovska, F. Hiroshi, T.i Ichiro, V. Nagarajan. Enhancement of dielectric properties in epitaxial bismuth ferrite – bismuth samarium ferrite superlattices**.** Advanced Electronic Materials. (2016)**,** https://doi.org/10.1002/aelm.201600170

[8] P. Sharma, A. N. Morozovska, E. A. Eliseev, Qi Zhang, D. Sando, N. Valanoor, and J. Seidel. Specific Conductivity of a Ferroelectric Domain Wall. ACS Appl. Electron. Mater. **4**, 6, 2739–2746. (2022) (https://pubs.acs.org/doi/10.1021/acsaelm.2c00261)

[9] P.S. Sankara Rama Krishnan, A. N. Morozovska, E. A. Eliseev, Q. M. Ramasse, D. Kepaptsoglou, W.-I. Liang, Y.-H. Chu, P. Munroe and V. Nagarajan. Misfit strain driven cation inter-diffusion across an epitaxial multiferroic thin film interface. Journal of Applied Physics, **115**, 054103 (2014), https://doi.org/10.1063/1.4862556

[10] J. Zhou, H.-H. Huang, S. Kobayashi, S. Yasui, Ke Wang, E. Eliseev, A. Morozovska, Pu Yu; I. Takeuchi, Z. Hong, D. Sando, Qi Zhang, N. Valanoor. An Emergent Quadruple Phase Ensemble in doped Bismuth Ferrite Thin Films through Site and Strain Engineering. Advanced

Functional Materials, 2403410 (2024) https://doi.org/10.1002/adfm.202403410

[11] A. N. Morozovska, E. A. Eliseev, P.S.Sankara Rama Krishnan, A. Tselev, E. Strelkov, A. Borisevich, O. V. Varenyk, N. V. Morozovsky, P. Munroe, S. V. Kalinin and V. Nagarajan. Defect thermodynamics and kinetics in thin strained ferroelectric films: the interplay of possible mechanisms. Phys.Rev.B **89**, 054102 (2014), https://doi.org/10.1103/PhysRevB.89.054102

[12] T. K. Lin, C. Y. Shen, C. C. Kao, C. F. Chang, H. W. Chang, C. R. Wang, and C. S. Tu. Structural evolution, ferroelectric, and nanomechanical properties of $Bi_{1-x}Sm_xFeO_3$ films (x= 0.05–0.16) on glass substrates. Journal of Alloys and Compounds, **787**, 397 (2019), https://doi.org/10.1016/j.jallcom.2019.02.008

[13] A. Raghavan, R. Pant, I. Takeuchi, E. A. Eliseev, M. Checa, A. N. Morozovska, M. Ziatdinov, S. V. Kalinin, Y. Liu, Evolution of ferroelectric properties in $Sm_xBi_{1-x}FeO_3$ via automated Piezoresponse Force Microscopy across combinatorial spread libraries, (2024), https://doi.org/10.1021/acsnano.4c06380

[14] G.B. Stephenson and M.J. Highland, Equilibrium and stability of polarization in ultrathin ferroelectric films with ionic surface compensation. Physical Review **B**, **84** (6), 064107, (2011) https://doi.org/10.1103/PhysRevB.84.064107

[15] M. J. Highland, T. T. Fister, D. D. Fong, P. H. Fuoss, Carol Thompson, J. A. Eastman, S. K. Streiffer, and G. B. Stephenson, Equilibrium polarization of ultrathin PbTiO3 with surface compensation controlled by oxygen partial pressure, Physical Review Letters,**107**, no. 18, (2011) 187602.

[16] R.V. Wang, D.D. Fong, F. Jiang, M.J. Highland, P.H. Fuoss, C. Tompson, A.M. Kolpak, J.A. Eastman, S.K. Streiffer, A.M. Rappe, and G.B. Stephenson, Reversible chemical switching of a ferroelectric film, Phys. Rev. Lett. **102**, 047601 (2009).

[17] D.D. Fong, A.M. Kolpak, J.A. Eastman, S.K. Streiffer, P.H. Fuoss, G.B. Stephenson, C.Thompson, D.M. Kim, K.J. Choi, C.B. Eom, I. Grinberg, and A.M. Rappe. Stabilization of Monodomain Polarization in Ultrathin PbTiO3 Films. Phys. Rev. Lett. **96**, 127601 (2006).

[18] M.J. Highland, T.T. Fister, M.-I. Richard, D.D. Fong, P.H. Fuoss, C.Thompson, J.A. Eastman, S.K. Streiffer, and G.B.Stephenson. Polarization Switching without Domain Formation at the Intrinsic Coercive Field in Ultrathin Ferroelectric $PbTiO_3$. Phys. Rev. Lett. **105**, 167601 (2010).

[19] S. M. Yang, A. N. Morozovska, R. Kumar, E. A. Eliseev, Ye Cao, L. Mazet, N. Balke, S. Jesse, R. Vasudevan, C. Dubourdieu, S. V. Kalinin. Mixed electrochemical-ferroelectric states in nanoscale ferroelectrics. Nature Physics **13**, 812 (2017), https://doi.org/10.1038/nphys4103

[20] A. N. Morozovska, E. A. Eliseev, A. Biswas, H. V. Shevliakova, N. V. Morozovsky, and S. V. Kalinin. Chemical control of polarization in thin strained films of a multiaxial ferroelectric:

phase diagrams and polarization rotation. Phys.Rev.B, **105**, 094112 (2022), https://link.aps.org/doi/10.1103/PhysRevB.105.094112

[21] A. N. Morozovska, E. A. Eliseev, A. Biswas, N. V. Morozovsky, and S. V. Kalinin. Effect of surface ionic screening on polarization reversal and phase diagrams in thin antiferroelectric films for information and energy storage. Phys. Rev. Applied **16**, 044053 (2021), http://link.aps.org/doi/10.1103/PhysRevApplied.16.044053

[22] K. P. Kelley, A. N. Morozovska, E. A. Eliseev, Y. Liu, S. S. Fields, S. T. Jaszewski, T. Mimura, J. F. Ihlefeld, S. V. Kalinin. Ferroelectricity in Hafnia Controlled via Surface Electrochemical State. Nature Materials **22**, 1144 (2023), https://doi.org/10.1038/s41563-023-01619-9

[23] E. A. Eliseev, S.i V. Kalinin, A. N. Morozovska. Ferro-ionic States and Domains Morphology in $Hf_xZr_{1-x}O_2$ Nanoparticles, J.Appl.Phys. (2025), https://doi.org/10.48550/arXiv.2410.04476

[24] A. N. Morozovska, S. V. Kalinin, M. E. Yelisieiev, J. Yang, M. Ahmadi, E. A. Eliseev, and Dean R. Evans. Dynamic control of ferroionic states in ferroelectric nanoparticles. Acta Materialia **237**, 118138 (2022), https://doi.org/10.1016/j.actamat.2022.118138

[25] S. V. Kalinin, E. A. Eliseev, and A. N. Morozovska. Adsorption of ions from aqueous solutions by ferroelectric nanoparticles (2024), https://doi.org/10.48550/arxiv.2409.07318

[26] A. N. Morozovska, S. V. Kalinin, E. A. Eliseev, S. Kopyl, Y. M. Vysochanskii, and D. R. Evans. Ferri-ionic Coupling in $CuInP_2S_6$ Nanoflakes: Polarization States and Controllable Negative Capacitance. Physical Review Applied **22,** 034059 (2024), https://link.aps.org/doi/10.1103/PhysRevApplied.22.034059

[27] A. N. Morozovska, E. A. Eliseev, I. V. Fesych, Y. O. Zagorodniy, O. S. Pylypchuk, E. V. Leonenko, M. V. Rallev, A. D. Yaremkevych, L. Demchenko, S. V. Kalinin, and O. M. Fesenko. Reentrant polar phase induced by the ferro-ionic coupling in $Bi_{1-x}Sm_xFeO_3$ nanoparticles, Physical Review B, **110**, 224110 (2024), https://doi.org/10.1103/PhysRevB.110.224110

[28] O.S. Pylypchuk, V.O. Kolupaiev, I.V. Fesych, V.N. Poroshin, A.N. Morozovska. Investigation of Electrophysical Properties, Phase Diagrams and Charge Carrier Transfer in $Bi_{1-x}Sm_xFeO_3$. Ukrainian Journal of Physics. **70** (10), 717, (2025); https://doi.org/10.15407/ujpe70.10.717

[29] A. N. Morozovska, E. A. Eliseev, N. V. Morozovsky, and S. V. Kalinin. Ferroionic states in ferroelectric thin films. Physical Review **B 95**, 195413 (2017); https://doi.org/10.1103/PhysRevB.95.195413)

[30] A. N. Morozovska, E. A. Eliseev, A. I. Kurchak, N. V. Morozovsky, R. K. Vasudevan, M. V. Strikha and S. V. Kalinin. Effect of surface ionic screening on polarization reversal scenario in

ferroelectric thin films: crossover from ferroionic to antiferroionic states. Physical Review **B** 96, 245405 (2017) https://doi.org/10.1103/PhysRevB.96.245405

[31] M.J. Highland, T.T. Fister, M.-I. Richard, D.D. Fong, P.H. Fuoss, C.Thompson, J.A. Eastman, S.K. Streiffer, and G.B.Stephenson. Polarization Switching without Domain Formation at the Intrinsic Coercive Field in Ultrathin Ferroelectric $PbTiO_3$. Phys. Rev. Lett. **105**, 167601 (2010).

[32] Victoria V. Khist, Eugene A. Eliseev, Maya D. Glinchuk, Dmitry V. Karpinsky, Maxim V. Silibin, and Anna N. Morozovska. Size Effects of Ferroelectric and Magnetoelectric Properties of Semi-ellipsoidal Bismuth Ferrite Nanoparticles. Journal of Alloys and Compounds, **714**, 15, 303–310 (2017) https://doi.org/10.1016/j.jallcom.2017.04.201

[33] E. A. Eliseev, V. V. Khist, Y. M. Fomichov, M. V. Silibin, G. S. Svechnikov, A. L. Kholkin, D. V. Karpinsky, V. V. Shvartsman, and A. N. Morozovska. Fixed Volume Effect on Polar Properties and Phase Diagrams of Ferroelectric Semi-ellipsoidal Nanoparticles. Eur. Phys. J. **B** 91: 150 (2018).
https://doi.org/10.1140/epjb/e2018-90133-6

[34] Anna N. Morozovska, Eugene A. Eliseev, Maya D. Glinchuk, Olena M. Fesenko, Vladimir V. Shvartsman, Venkatraman Gopalan, Maxim V. Silibin, Dmitry V. Karpinsky. Rotomagnetic coupling in fine-grained multiferroic $BiFeO_3$: theory and experiment. Phys.Rev. **B 97**, 134115 (2018) https://link.aps.org/doi/10.1103/PhysRevB.97.134115

[35] E. Palaimiene, J. Macutkevic, D. V. Karpinsky, A. L. Kholkin, and J. Banys. Dielectric investigations of polycrystalline samarium bismuth ferrite ceramic, Applied Physics Letters **106**, 012906 (2015); https://doi.org/10.1063/1.4905344

[36] D. V. Karpinsky, O. M. Fesenko, M. V. Silibin, S. V. Dubkov, M. Chaika, A. Yaremkevich, A. Lukowiak, Y. Gerasymchuk, W. Strek, A. Pakalniskis, R. Skaudzius, A. Kareiva, Y. M. Fomichov, V. V. Shvartsman, S. V. Kalinin, N. V. Morozovsky, and A. N. Morozovska. Ferromagnetic-like behavior of $Bi_{0.9}La_{0.1}FeO_3$-KBr nanocomposites. Scientific Reports **9**, Article number: 10417 (2019) https://www.nature.com/articles/s41598-019-46834-0

[37] A. Bougoffa, E. M. Benali, A. Benali, M. Bejar, E. Dhahri, M. P. Graça, M. A. Valente, G. Otero-Irurueta, B. F. Costa. Investigation of temperature and frequency dependence of the dielectric properties of multiferroic $(La_{0.8}Ca_{0.2})_{0.4}Bi_{0.6}FeO_3$ nanoparticles for energy storage application. RSC Advances **12, (11)** 6907–6917(2022); https://doi.org/10.1039/d1ra08975g

[38] K. Papadopoulos, E. Myrovali, D. Karfaridis, M. Farle, U.Wiedwald, M. Angelakeris. Cation substitution and size confinement effects on structure, magnetism and magnetic hyperthermia of $BiFeO_3$-based multiferroic nanoparticles and Hydrogels. Journal of Alloys and Compounds. **969**, 172337 (2023); https://doi.org/10.1016/j.jallcom.2023.172337

[39] H. Han, D. Ghosh, J.L. Jones, & J. C. Nino, Colossal Permittivity in Microwave-Sintered Barium Titanate and Effect of Annealing on Dielectric Properties. J. Am. Ceram. Soc., 96, 485 (2013). https://doi.org/10.1111/jace.12051

[40] J. Petzelt, D. Nuzhnyy, V. Bovtun, M. Savinov, M. Kempa, I. Rychetsky, Broadband dielectric and conductivity spectroscopy of inhomogeneous and composite conductors. Phys. Stat. Sol. A 210, 2259 (2013), https://doi.org/10.1002/pssa.201329288

[41] O. S. Pylypchuk, S. E. Ivanchenko, M. Y. Yelisieiev, A. S. Nikolenko, V. I. Styopkin, B. Pokhylko, V. Kushnir, D. O. Stetsenko, O. Bereznykov, O. V. Leschenko, E. A. Eliseev, V. N. Poroshin, N. V. Morozovsky, V. V. Vainberg, and A. N. Morozovska. Behavior of the Dielectric and Pyroelectric Responses of Ferroelectric Fine-Grained Ceramics. Journal of the American Ceramic Society, **108**, e20391 (2025); https://doi.org/10.1111/jace.20391.

[42] O. S. Pylypchuk, V. V. Vainberg, V. N. Poroshin, O. V. Leshchenko, V. N. Pavlikov, I. V. Kondakova, S. E. Ivanchenko, L. P. Yurchenko, L. Demchenko, A. O. Diachenko, M. V. Karpets, M. P. Trubitsyn, E. A. Eliseev, and A. N. Morozovska. A colossal dielectric response of $Hf_xZr_{1-x}O_2$ nanoparticles. Physical Review Materials **9**, 114412 (2025); https://doi.org/10.1103/y2pb-5g5w

[43] M. Saleem, M. S. Butt, A. Maqbool, M. A. Umer, M. Shahid, F. Javaid, R. A. Malik, H. Jabbar, H. M. W. Khalil, L. D. Hwan, M. Kim, B.-K. Koo, S. J. Jeong, "Percolation phenomena of dielectric permittivity of a microwave-sintered $BaTiO_3$-Ag nanocomposite for high energy capacitor", Journal of Alloys and Compounds, **822**, 153525 (2020), https://doi.org/10.1016/j.jallcom.2019.153525

[44] P. Lunkenheimer, S. Krohns, S. Riegg, S.G. Ebbinghaus, A. Reller, and A. Loidl, Colossal dielectric constants in transition-metal oxides. Eur. Phys. J. Spec. Topics, *180*, 61-89 (2009), https://doi.org/10.1140/epjst/e2010-01212-5

[45] J. Petzelt, D. Nuzhnyy, V. Bovtun, D.A. Crandles. Origin of the colossal permittivity of (Nb+ In) co-doped rutile ceramics by wide-range dielectric spectroscopy. Phase Transitions **91**, 932 (2018). https://doi.org/10.1080/01411594.2018.1501801

[46] I. Rychetský, D. Nuzhnyy, J. Petzelt, Giant permittivity effects from the core–shell structure modeling of the dielectric spectra. Ferroelectrics **569**, 9 (2020). https://doi.org/10.1080/00150193.2020.1791659

[47] A. N. Morozovska, E. A. Eliseev, D. Chen, C. T. Nelson, and S. V. Kalinin. Building Free Energy Functional from Atomically-Resolved Imaging: Atomic Scale Phenomena in La-doped $BiFeO_3$. Phys.Rev. B, **99**, 195440 (2019); https://link.aps.org/doi/10.1103/PhysRevB.99.195440

[48] W. Heywang. Semiconducting Barium Titanate. J. Materials Science **6**, 1214 (1971); https://doi.org/10.1007/BF00550094